%% file: main.tex
\documentclass[sigconf]{acmart}
\AtBeginDocument{%
  }

\usepackage{soul}

\usepackage{xspace}
\usepackage{amsmath}
\usepackage{tcolorbox}
\usepackage{listings}
\usepackage{xcolor}
\usepackage{fancyvrb}
\usepackage{fancyvrb} 
\usepackage{tabularx}
\usepackage{algpseudocode}
\usepackage{algorithm}

\usepackage{caption}
\usepackage{float}
\usepackage{graphicx}

\newcommand{\tool}{\textit{LLM4FL}\xspace}
\newcommand{\tooltable}{LLM4FL\xspace}

\newcommand{\toolrandom}{\textit{LLM4FL\textsubscript{Execution}}\xspace}
\newcommand{\toolrandomtable}{LLM4FL\textsubscript{Execution}\xspace}

\newcommand{\toolochiai}{\textit{LLM4FL\textsubscript{Ochiai}}\xspace}
\newcommand{\toolochiaitable}{LLM4FL\textsubscript{Ochiai}\xspace}

\newcommand{\tooldepgraph}{\textit{LLM4FL\textsubscript{DepGraph}}\xspace}
\newcommand{\tooldepgraphtable}{LLM4FL\textsubscript{DepGraph}\xspace}

\newcommand{\toolwopromptchain}{\textit{LLM4FL\textsubscript{w/o~CodeNav}}\xspace}
\newcommand{\toolwopromptchaintable}{LLM4FL\textsubscript{w/o~CodeNav}\xspace}

\newcommand{\toolwosplitting}{\textit{LLM4FL\textsubscript{w/o~Division}}\xspace}
\newcommand{\toolwosplittingtable}{LLM4FL\textsubscript{w/o~Division}\xspace}

\newcommand{\toolworeflection}{\textit{LLM4FL\textsubscript{w/o~Reflexion}}\xspace}
\newcommand{\toolworeflectiontable}{LLM4FL\textsubscript{w/o~Reflexion}\xspace}

\newcommand{\ochiai}{\textit{Ochiai}\xspace}
\newcommand{\Grace}{\textit{Grace}\xspace}
\newcommand{\depgraph}{\textit{DepGraph}\xspace}
\newcommand{\deepfl}{\textit{DeepFL}\xspace}

\usepackage[utf8]{inputenc}

\definecolor{deepgreen}{rgb}{0.0, 0.5, 0.0}
\definecolor{ForestGreen}{rgb}{0.0, 0.5, 0.0}

\newcommand{\phead}[1]{\noindent {\bf #1}}
\newcommand{\uhead}[1]{\textit{#1}}

\newcommand{\rqboxc}[1]{\begin{tcolorbox}[left=4pt,right=4pt,top=4pt,bottom=4pt,colback=gray!35,colframe=gray!35,before skip=3pt,after skip=3pt]#1\end{tcolorbox}}

\definecolor{lightgray}{gray}{0.9}

\newcommand\greybox[1]{%
  \vskip\baselineskip%
  \vspace{-0.8\baselineskip}
  \par\noindent\colorbox{lightgray}{%
    \begin{minipage}{\linewidth}#1\end{minipage}%
  }%
  \vspace{-0.8\baselineskip}
  \vskip\baselineskip%
}

\setcopyright{acmlicensed}
\copyrightyear{2018}
\acmYear{2018}
\acmDOI{XXXXXXX.XXXXXXX}
\acmConference[Conference acronym 'XX]{Make sure to enter the correct
  conference title from your rights confirmation email}{June 03--05,
  2018}{Woodstock, NY}
\acmISBN{978-1-4503-XXXX-X/2018/06}

\begin{document}


\title{A Multi-Agent Approach to Fault Localization via Graph-Based Retrieval and Reflexion}

\author{Md Nakhla Rafi}
\orcid{0009-0005-4707-8985}
\affiliation{%
  \institution{Software Performance, Analysis, \\and Reliability (SPEAR) Lab\\Concordia University}
  \city{Montreal}
  \country{Canada}
}
\email{mdnakhla.rafi@mail.concordia.ca}

\author{Dong Jae Kim}
\orcid{0000-0002-3181-0001}
\affiliation{%
  \institution{DePaul University}
  \city{Chicago}
  \country{USA}
}
\email{dkim121@depaul.edu}

\author{Tse-Hsun (Peter) Chen}
\orcid{0000-0003-4027-0905}
\affiliation{%
  \institution{Software Performance, Analysis, \\and Reliability (SPEAR) Lab\\Concordia University}
  \city{Montreal}
  \country{Canada}
}
\email{peterc@encs.concordia.ca}

\author{Shaowei Wang}
\orcid{0000-0003-3823-1771}
\affiliation{%
  \institution{University of Manitoba}
  \city{Winnipeg}
  \country{Canada}
}
\email{Shaowei.Wang@umanitoba.ca}

\renewcommand{\shortauthors}{Trovato et al.}

\begin{abstract}
Identifying and resolving software faults remains a challenging and resource-intensive process. Traditional fault localization techniques, such as Spectrum-Based Fault Localization (SBFL), leverage statistical analysis of test coverage but often suffer from limited accuracy. While learning-based approaches improve fault localization, they demand extensive training datasets and high computational resources. Recent advances in Large Language Models (LLMs) offer new opportunities by enhancing code understanding and reasoning. However, existing LLM-based fault localization techniques face significant challenges, including token limitations, performance degradation with long inputs, and scalability issues in complex software systems. To overcome these obstacles, we propose \textit{LLM4FL}, a multi-agent fault localization framework that utilizes three specialized LLM agents. First, the \textit{Context Extraction Agent} applies an order-sensitive segmentation strategy to partition large coverage data within the LLM’s token limit, analyze failure context, and prioritize failure-related methods. The extracted data is then processed by the \textit{Debugger Agent}, which employs graph-based retrieval-augmented code navigation to reason about failure causes and rank suspicious methods. Finally, the \textit{Reviewer Agent} re-evaluates the identified faulty methods using verbal reinforcement learning, engaging in self-criticism and iterative refinement. Evaluated on the Defects4J (V2.0.0) benchmark, which includes 675 faults from 14 Java projects, \textit{LLM4FL} achieves an 18.55\% improvement in Top-1 accuracy over \textit{AutoFL} and 4.82\% over \textit{SoapFL}. It outperforms supervised techniques such as \textit{DeepFL} and \textit{Grace}, all without requiring task-specific training. Furthermore, its coverage segmentation and prompt chaining strategies enhance performance, increasing Top-1 accuracy by up to 22\%.
\end{abstract}

\ccsdesc[500]{Do Not Use This Code~Generate the Correct Terms for Your Paper}
\ccsdesc[300]{Do Not Use This Code~Generate the Correct Terms for Your Paper}
\ccsdesc{Do Not Use This Code~Generate the Correct Terms for Your Paper}
\ccsdesc[100]{Do Not Use This Code~Generate the Correct Terms for Your Paper}

\keywords{Fault localization, large language model, agent}


\maketitle
\input{samples/intro.tex}
\input{samples/background}

\input{samples/methodology}

\input{samples/results}
\input{samples/threats}
\input{samples/conclusion}

\balance
\bibliographystyle{ACM-Reference-Format}
\bibliography{reference}

\end{document}

%% file: samples/intro.tex
\section{Introduction}
The process of locating and fixing software faults requires significant time and effort. Research shows that software development teams allocate more than half of their budgets to testing and debugging activities~\cite{hait2002economic, alaboudi2021exploratory}. 
As software systems become increasingly complex, the demand for more accurate fault localization techniques grows. Hence, to assist developers and reduce debugging costs, researchers have developed various fault localization techniques~\cite{li2019deepfl, lou2021boosting, qian2023gnet4fl, li2021fault, abreu2009spectrum, sohn2017fluccs}. These techniques analyze code coverage and program execution to identify the most likely faulty code, assisting developers in finding the fault.

Despite the advances in fault localization, many existing techniques still struggle with scalability and precision. Traditional methods, such as Spectrum-Based Fault Localization (SBFL), use statistical analysis to analyze coverage data from passing and failing test cases to rank suspicious code elements~\cite{abreu2006evaluation}. While these techniques provide valuable insights, their accuracy is lower. Their reliance on statistical correlations between test failures and code coverage does not always capture the deeper semantic relationships needed for more accurate fault localization~\cite{wong2016survey,xie2013theoretical,le2013theory}. Recent techniques applied machine learning~\cite{sohn2017fluccs, zhang2019empirical, li2021fault, li2019deepfl} and deep learning models~\cite{lou2021boosting, qian2023gnet4fl, depgraph, lou2021boosting} to address these issues to improve fault localization. These methods enhance the ranking of suspicious code elements by incorporating additional information like code complexity, text similarity, and historical fault data. 
However, these techniques often require extensive training data that may not be available.

Recent advances in Large Language Models (LLMs) have shown great potential for software fault localization due to their strong language comprehension and generation capabilities~\cite{abedu2024llm, lin2024llm}. LLMs trained on extensive programming datasets can understand code structure, interpret error messages, and even suggest bug fixes~\cite{wu2023large, pu2023summarization, li2023explaining}. These models, with their ability to analyze and process both natural language and code, present an opportunity to significantly improve traditional fault localization methods by incorporating deeper semantic analysis and context-aware reasoning. 

However, there are several challenges in using LLMs for fault localization. Firstly, LLMs face significant difficulties when analyzing large-scale repository-level data, primarily due to inherent token limitations, as highlighted in recent studies~\cite{hadi2023large, hou2023large, wu2023large}. Secondly, the performance of LLMs tends to decrease when they are applied to complex systems that require reasoning about multiple interacting components. This complexity can negatively impact the performance of LLMs~\cite{liu2024lost, wu2023large}. LLMs may struggle to understand the inter-procedural method call relationships, an essential aspect for understanding how faults propagate across the entire software repository~\cite{chatterjee2015quantitative,levy2024same, liu2024lost}.

In this paper, we propose \tool, a multi-agent LLM-based fault localization technique designed to analyze large-scale software projects at the repository level.
\tool leverages a multi-agent framework that emulates developers' fault localization process. \tool employs three LLM agents, each with a specialized role to enhance fault localization effectiveness. The \textit{\textbf{Context Extraction Agent}} divides the methods covered by the failing test cases into smaller groups, each within LLM's token limitation. 
Then, the agent analyzes the test code and test failure messages to reason about possible failure causes. This helps prioritize the most suspicious covered methods from each group before passing them on to the next agent.
Next, the \textit{\textbf{Debugger Agent}} analyzes each covered method and performs graph-based, retrieval-augmented (Graph-RAG) code navigation. It traverses the prioritized methods' code iteratively along call dependencies to analyze and rank suspicious methods. Finally, the \textit{\textbf{Reviewer Agent}} re-ranks the identified fault methods collected from all groups using verbal reinforcement learning~\cite{shinn2024reflexion}, criticizing and refining each response to ensure more precise fault localization.

We evaluated \tool using the Defects4J (V2.0.0) benchmark~\cite{just2014defects4j}, which contains 675 real-world faults from 14 open-source Java projects. Our results demonstrate that \tool surpasses LLM-based technique \textbf{\textit{AutoFL}}~\cite{autofl} and \textbf{\textit{AgentFL/SoapFL}}~\cite{qin2024agentfl} by achieving 18.55\% and 4.82\% higher Top-1 accuracy, respectively. \tool is also the cheapest among the three, costing only \$0.05 per fault. Additionally, \tool outperforms supervised techniques such as \deepfl~\cite{li2019deepfl} and \Grace~\cite{lou2021boosting}, even without task-specific training. We also analyzed the impact of individual components within \tool on fault localization accuracy. Our findings indicate that each component plays a significant role in performance. Among these, dividing the covered methods into smaller groups, and Graph-RAG-based code navigation contributes the most. 
Moreover, we examined whether the initial ordering of methods provided to the LLM influences performance. 
The results reveal that method ordering is important. Even though LLM eventually visits all methods, the  Top-1 accuracy still has a difference of up to 22\% when comparing an execution-based ordering and the order provided by \depgraph~\cite{depgraph}. 

The paper makes the following contributions:
\begin{itemize}
     \item We introduce \tool, a novel LLM-based fault localization technique that employs a divide-and-conquer strategy. This technique groups large coverage data and ranks the covered methods using an SBFL formula. Using multiple agents and code navigation based on Graph-RAG, \tool analyzes the repository iteratively to identify and localize faults.

    \item We conducted an extensive evaluation and \tool demonstrates superior performance, surpassing \textit{AutoFL}~\cite{autofl} by 18.55\% and \textit{AgentFL/SoapFL}~\cite{qin2024agentfl} by 4.82\% in Top-1 accuracy. It also outperforms supervised techniques like \textit{DeepFL} and \textit{Grace}, achieving these results without requiring task-specific training.

    \item Our analysis of \tool's components shows that key components like dividing methods into smaller groups and code navigation are essential to its fault localization accuracy. Removing these features leads to performance declines, emphasizing their importance in handling token limitations and effective fault analysis.
    
    \item We find that 
    the different ordering on the initial method list passed to LLM can affect fault localization accuracy by up to 22\% in Top-1 scores, even though LLM eventually visits all the methods. 

\end{itemize}

In short, we provide a strategy to mitigate the token limitation issues and analyze repository-level data in LLM-based fault localization. We also highlight the impact of initial method ordering for LLM's input. The findings may help inspire future research on LLM-based fault localization for large-scale software projects. 

\noindent {\bf Paper Organization.} Section \ref{background} discusses background and related work. Section \ref{Methodology} describes our technique, \tool. Section \ref{studyrq} presents the experimental results. Section \ref{threats} discusses the threats to validity. Section \ref{conclusion} concludes the paper. 

%% file: samples/background.tex
\begin{table*}[h]
\centering
\caption{Comparison of LLM-based fault localization techniques.}
\scalebox{0.8}{
\begin{tabular}{p{1.8cm}|p{4.5cm}|p{4.5cm}|p{4.5cm}|p{4.5cm}}
    \toprule
    \textbf{Aspect} & \textbf{LLM4FL (this work)} & \textbf{AutoFL~\cite{autofl}} & \textbf{SoapFL/AgentFL~\cite{qin2024agentfl}} & \textbf{AutoCodeRover \cite{zhang2024autocoderover}} \\
    \midrule
    \textbf{Navigating large code repository} & The agent autonomously navigates the repository, iteratively following an inter-procedural call graph to retrieve and explore connected methods.
    & Retrieve class and method signatures related to the failure, then let the LLM decide which methods need deeper analysis. & Analyze the test failures to identify the most suspicious classes, then iteratively analyze related methods. & Start with the file or method names extracted from issue reports, then iteratively search the repository based on called methods in the retrieved source code. \\
    \midrule
    \textbf{Handling token limitation} & Divides methods into groups and analyzes each group separately. & Provide only class and method signatures to the LLM, then let it decide which methods need deeper analysis. & Use LLMs to generate description for classes and methods using their signatures. Then, the LLM decides which classes/methods are relevant to the test failures for further analysis. 
    & Iteratively search for relevant classes and methods based on the issue description and prioritize methods ranked by SBFL techniques. \\
    \midrule
    \textbf{Ranking of faulty methods} & Hierarchical ranking. First, it finds the suspicious methods and ranks them in each group. Then, it employs verbal reinforcement learning to combine and re-rank the suspicious methods from all groups.  & Re-run the fault localization process multiple times and do a majority vote on the result. &  Utilize a multi-round dialogue strategy that allows LLM agents to discuss the code to score and rank faulty methods. & Does not rank methods; it selects a method as faulty from the retrieved context and generates a patch.\\
    \bottomrule
\end{tabular}}
\label{tab:llm_comparison}
\end{table*}

\section{Background and Related Work}
\label{background}
\subsection{Background}
\phead{Large Language Models.} Large Language Models (LLMs), primarily built on the transformer architecture~\cite{meta2024llama3, brown2020language, roziere2023code}, have significantly advanced the field of natural language processing (NLP). These LLMs, such as the widely recognized GPT3 model with its 175 billion parameters~\cite{brown2020language}, are trained on diverse text data from various sources, including source code. The training involves self-supervised learning objectives that enable these models to develop a deep understanding of language and generate contextually relevant and semantically coherent text. LLMs have shown substantial capability in tasks that involve complex language comprehension and generation~\cite{abedu2024llm, lin2024llm}, such as code recognition and generation. Recent research has leveraged LLMs in software engineering tasks, particularly in fault localization~\cite{autofl, qin2024agentfl, testfreefaultlocalization}, where they assist in identifying the faulty code groups responsible for software errors. One of the key advantages of using LLMs in fault localization is their ability to process both natural language and code without re-training, allowing them to analyze error messages, stack traces, and test case information to infer suspicious methods or code sections in an unsupervised zero-shot setting.

\phead{LLM Agents.} LLM agents leverage LLMs to autonomously execute tasks described in natural language, making them versatile tools across various domains.
LLM agents are artificial intelligence systems that utilize LLMs as their core computational engines to understand questions and generate human-like responses. They leverage functionalities like memory management~\cite{zhou2023recurrentgpt} and tool integration~\cite{xia2024agentless, roy2024exploring} to handle multi-step and complex operations seamlessly. 
The agents can refine their responses based on feedback, learn from new information, and even interact with other AI agents to collaboratively solve complex tasks~\cite{hong2024metagpt, qian2023communicative, xu2023exploring, lin2024llm}. Through prompting, agents can be assigned different roles (e.g., a developer or a tester), providing more domain-specific responses that help improve the answer~\cite{hong2024metagpt,white2024chatgpt, shao2023character}. 

Recent studies~\cite{shinn2024reflexion,renze2024self,pan2025codecor} explore using the agent's verbal reasoning result to guide iterative improvement (i.e., \textbf{verbal reinforcement learning}), which has shown promising improvement in downstream tasks. 
In verbal reinforcement learning, LLM agents receive natural language feedback, such as reasoning results or instructions, from other agents as a reward signal. This allows agents to learn and adapt their behavior based on human-like guidance, improving their learning process to solve complex tasks. 
As the capabilities of large language model (LLM) agents grow, they play an essential role in enhancing automation and increasing productivity in software development. They can assist in code generation~\cite{nijkamp2022codegen,lin2024soen,gu2023llm}, debugging~\cite{lee2024unified,autofl}, test case creation~\cite{huang2024generative,chen2024chatunitest}, and automated refactoring~\cite{pomian2024next,liu2025exploring}, enabling developers to streamline repetitive tasks and focus on higher-level design and problem-solving. Additionally, LLM agents can facilitate collaborative software engineering by acting as intelligent assistants in code reviews, documentation generation, and issue resolution, improving overall development efficiency~\cite{lin2024soen,he2024llm}. 
In this paper, we explore using LLM agents to improve fault localization by emulating developers' debugging process using verbal reinforcement learning.

\subsection{Related Work}

\phead{Spectrum-based Fault Localization.} Spectrum-based fault localization (SBFL)~\cite{abreu2006evaluation,jones2002visualization,wong2013dstar,abreu2009spectrum} employs statistical techniques to evaluate the suspiciousness of individual code elements, such as methods, by analyzing test outcomes and execution traces. The core idea of SBFL is that code components that are executed more frequently in failing tests and less frequently in passing tests are more likely to contain faults. Despite its widespread study, SBFL's practical effectiveness remains limited~\cite{kochhar2016practitioners,xie2016revisit}. To enhance SBFL's accuracy, recent research~\cite{cui2020improving, wen2019historical, chen2022useful, xu2020every} has suggested incorporating additional data, such as code changes~\cite{wen2019historical,chen2022useful} or mutation analysis~\cite{cui2020improving,xu2020every}. However, SBFL's reliance on code coverage metrics still poses challenges, as its suspiciousness scores may not generalize effectively to different faults or systems.

\phead{Learning-based fault localization.}
Recent efforts have focused on improving SBFL with learning-based methods~\cite{sohn2017fluccs, zhang2019empirical, li2021fault, li2017transforming, li2019deepfl, zhang2019cnn}. These approaches use machine learning models like radial basis function networks~\cite{wong2011effective}, back-propagation networks~\cite{wong2009bp}, and convolutional neural networks~\cite{zhang2019cnn, li2021fault, albawi2017understanding} to estimate suspiciousness scores based on historical faults. Some techniques, such as \textit{FLUCCS}~\cite{sohn2017fluccs}, combine SBFL scores with metrics like code complexity, while others, like \textit{DeepFL}~\cite{li2019deepfl} and \textit{CombineFL}~\cite{zou2019empirical}, merge multiple sources such as spectrum-based and mutation-based data~\cite{moon2014ask, papadakis2015metallaxis, dutta2021msfl}. Graph neural networks (GNNs) have also been applied to fault localization~\cite{qian2023gnet4fl, lou2021boosting, qian2021agfl, xu2020defect}. Techniques like \textit{Grace}~\cite{lou2021boosting} and \textit{GNET4FL}~\cite{qian2023gnet4fl} utilize test coverage and source code structure for improved accuracy, while \textit{DepGraph}~\cite{depgraph} refines these approaches by graph pruning and incorporating code change information, resulting in higher performance with reduced computational demands. Although these learning-based techniques show improved results, they require training data that may not be available to every project. 

\phead{LLM-Based Fault Localization.} Large Language Models (LLMs), such as GPT-4o~\cite{openai_gpt4o_2024}, LLaMA~\cite{meta2024llama3}, and ChatGPT~\cite{achiam2023gpt}, demonstrated remarkable abilities in processing both natural and programming languages. LLMs 
have shown potential in identifying and fixing errors using program code and error logs~\cite{achiam2023gpt}. 
Some LLM-based fault localization techniques operate on small code snippets due to LLM's token limitations. \textit{LLMAO}~\cite{testfreefaultlocalization} uses bidirectional adapters to score suspicious lines within a 128-line context. In contrast, Wu et al.~\cite{wu2023large} prompt ChatGPT with code and error logs but struggle to scale to large projects~\cite{liu2024lost}.

Recent fault localization techniques employ different strategies to navigate large codebases and refine fault localization results~\cite{autofl,zhang2024autocoderover,qin2024agentfl}. Each technique consists of key components for retrieving failure-related classes or methods, navigating the code repository, and ranking suspicious methods. However, there are differences in how these techniques design the components. Table \ref{tab:llm_comparison} provides a summary of the key differences between prior techniques and ours.
\textit{AutoFL}~\cite{autofl} retrieves class and method signatures related to the test failure, and prompts the LLM to decide which methods need future analysis (i.e., to examine the code).  
While this approach ensures the input size is less than the LLM's token limitation, it does not consider method dependence in the analysis. Instead, it refines rankings by repeatedly running the fault localization process and using a majority voting mechanism to determine the most likely faulty methods. 
\textit{AgentFL/SoapFL}~\cite{qin2024agentfl} models fault localization as a structured operating procedure with controlled phases. It uses a document-guided approach, where the system creates (or refines) LLM-generated or developer-provided documentation for classes and methods, then navigates the codebase based on these documents. By iteratively scoring and ranking suspicious methods, it continuously narrows down the fault candidates. It integrates document-guided search and enhanced method documentation to navigate the codebase while using iterative LLM-based scoring and ranking to refine the set of suspicious methods.
\textit{AutoCodeRover}~\cite{zhang2024autocoderover} first extracts file/method names from an issue report. Then, based on the names, it searches for corresponding classes and methods. It is an iterative search process and may continue to search for  the called methods in the repository based on LLM feedback. 
\textit{AutoCodeRover} may also be integrated with SBFL to refine localization and guide patch generation. However, it does not explicitly rank faulty methods; instead, it selects a method from the retrieved context and generates a patch. 


In contrast to prior techniques, \tool uses a structured, graph-based approach to analyze the code, enabling the LLM to systematically navigate the repository by considering caller-callee relationships. This eliminates LLM hallucination when searching and retrieving methods in the repository.  
Additionally, \tool uses a divide-and-conquer strategy to mitigate token limitations by breaking down large methods into smaller, manageable groups, ensuring comprehensive analysis. Finally, \tool refines fault localization results through verbal reinforcement learning, iteratively re-ranking suspicious methods for improved accuracy. We believe these contributions collectively enhance the accuracy and scalability of LLM-based fault localization.

%% file: samples/methodology.tex
\section{Methodology}
\label{Methodology}
Traditionally, developers rely on dynamic execution data to facilitate fault localization: (i) code coverage is used for narrowing down executed code that changes the behavior incorrectly, whereas (ii) stack trace provides developers with insight into the program execution when the exception gets thrown, helping them diagnose the issue more effectively. 
However, directly applying Large Language Models (LLMs) for fault localization is infeasible. This is due to the large size of software artifacts, which frequently exceed token limitations, leading to input truncation and degraded performance ~\cite{levy2024same, liu2024lost, wu2023large}. To mitigate this issue, we propose \tool, an LLM-agent-based fault localization system, which adopts a divide-and-conquer algorithm, which first divides the large software artifacts, and employs multiple LLM-Agents to localize faults autonomously.

\begin{figure*}[ht]
    \centering
    \includegraphics[width=0.9\textwidth]
    {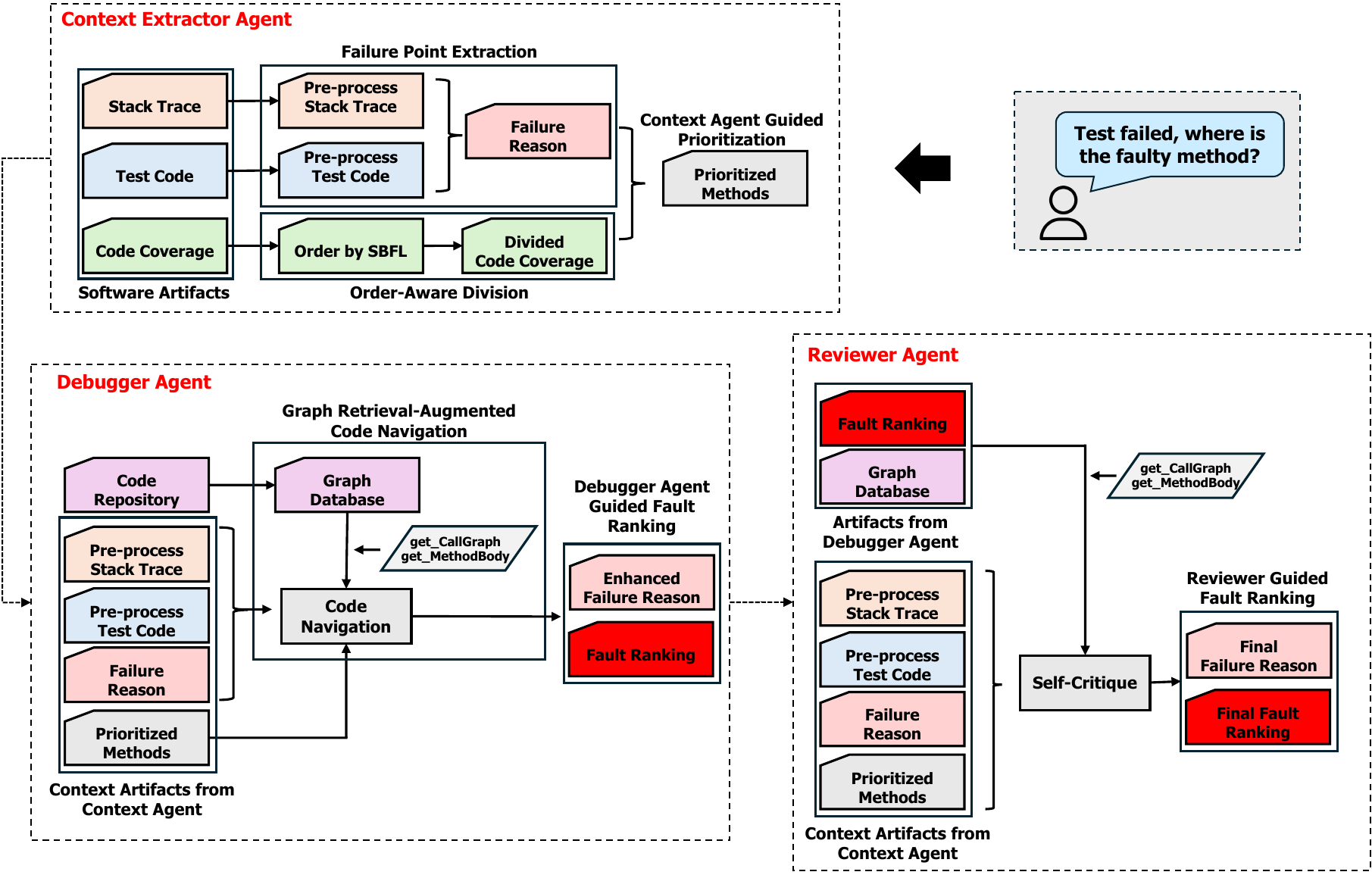}
    \caption{Overview of \tool for Multi-Agent Fault Localization, illustrating how agents collaborate to analyze software artifacts, extract failure reasoning, perform graph-based retrieval-augmented code navigation, and rank faulty methods using verbal reinforcement learning.}
    \label{fig:overall-approach}
\end{figure*}

Figure~\ref{fig:overall-approach} provides an overview of our technique. \tool consists of three LLM-Agents by using novel prompting techniques: (i) \textit{Context Extraction Agent}, (ii) \textit{Debugger Agent}, and (iii) \textit{Reviewer Agent} to localize the fault iteratively. Specifically, the \textit{Context Extraction Agent} utilizes a \textbf{order-aware division} and \textbf{failure-reason guided prioritization}
prompting technique, which consists of two phases. In the {division phase}, a 
toolchain divides the large-scale code coverage data into small groups to fit within the LLM’s token limits. In the {prioritization phase}, the agent iteratively analyzes the divided code coverage, leveraging failed test cases and stack traces to identify potentially faulty methods. The \textit{Debugger Agent} then performs \textbf{graph-based retrieval-augmented code navigation} prompting to locate code artifacts further to enhance the accuracy of fault ranking. Finally, the \textit{Reviewer Agent} re-ranks the buggy method through \textbf{verbal reinforcement learning} prompting with the reviewer agent, ensuring more precise fault localization. 
Below, we provide more details on each of these phases.

\subsection{Context Extraction Agent}~\label{sec:splitting}
This agent defines a \textbf{order-aware division} and \textbf{failure-reason guided prioritization} prompting technique to divide code-coverage and iteratively prioritize faulty methods using stack trace and test code within each division. We describe the technique below.
\subsubsection{Order-Aware Division Phase}
To mitigate token size limitation, our \tool first performs a division phase, where it first runs the test case to extract the list of method-level code coverage using GZoltar~\cite{campos2012gzoltar}, which is denoted as \(C\), containing a sequence of pairs (\(m\), \(s\)), where \(m\) denotes the method and \(s\) denotes the set of statements within method \(m\). We then divide \(C\) into $K$ group of sequences of \( C_1, C_2, \ldots, C_k \), where \( K = \left\lceil \frac{\text{Input Token Length}}{\text{Token Limitation}} \right\rceil \), and each subset satisfies \( |C_i| \leq \lfloor \text{Token Limitation} \rfloor \) to ensure the data fits within the LLM's context window. For example, given an \textit{Input Token Length} of 500K tokens, we divide this by the \textit{Token Limitation} for \textsf{GPT-4o-mini}, which has a limit of 128K tokens~\cite{openai_gpt4o_2024}. This results in \( K = 4 \), each with 128K token limit.

Traditionally, the order in which divisions occur in divide-and-conquer algorithms does not impact the outcome. However, prior studies have shown that LLM may improve performance when the order of instructions is carefully considered ~\cite{chen2024premise}. Inspired by this, we propose \textit{order-aware division}, where we use a Spectrum-Based Fault Localization (SBFL) technique to sort \(C\) based on their likelihood of being faulty. Specifically, we use \ochiai ranking to order \( C_1, C_2, \ldots, C_k \), which is an efficient and unsupervised technique that assigns higher suspiciousness scores to statements that are executed more frequently by failing test cases and less frequently by passing ones~\cite{abreu2006evaluation, lou2021boosting, li2021fault, cui2020improving, wen2019historical, qian2021agfl}. 

\subsubsection{Failure-Reason Guided Prioritization Phase} To prioritize faulty methods from the divided code coverage, one approach is to provide the agent with the stack trace and test case. However, prior studies have shown that incorporating summarized description can improve accuracy~\cite{stiennon2020learning, roit2023factually}. Thus, we propose failure-reason-guided fault prioritization, which follows two steps. First, the input tokens are summarized into a failure-reason representation, capturing the test purpose, expected output, and failure reason. Second, the divided code coverage is prioritized based on the reason for failure. Below, we discuss each step in detail.

\begin{figure}[ht]
    \centering
    \includegraphics[width=0.5\textwidth]
    {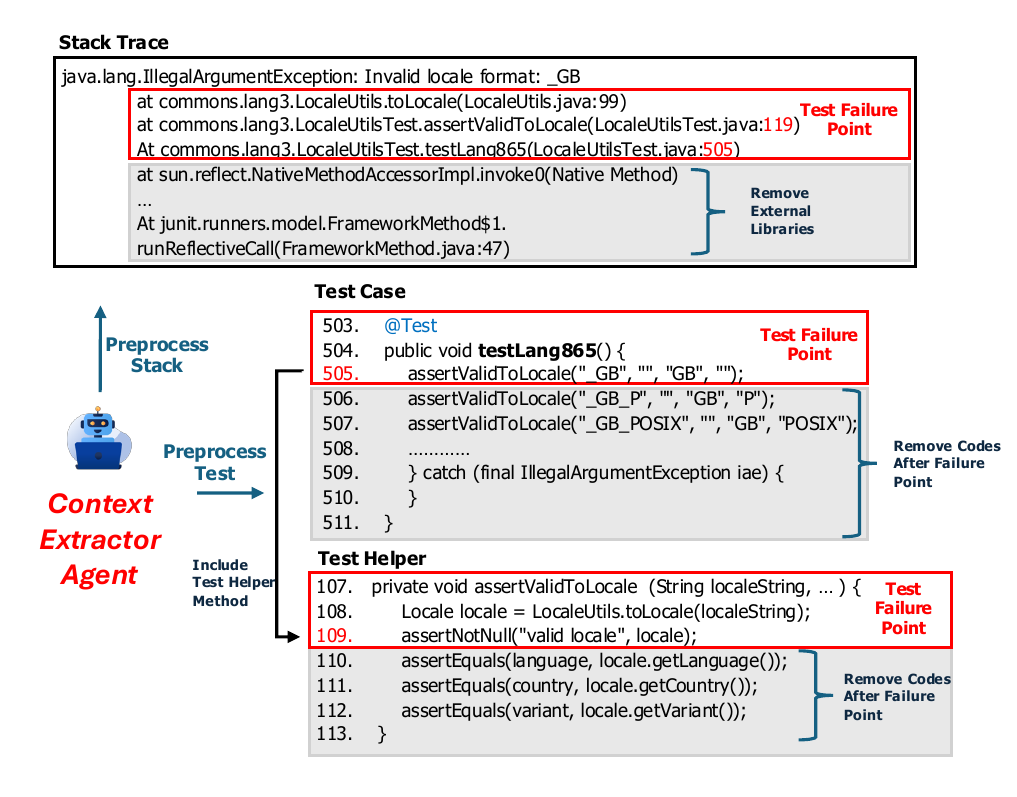}
    \caption{Context-Extraction Agent uses tool-chains to preprocess software artifacts for Lang-5 to emphasize the test failure context. For (i) stack-trace the agent prunes external libraries and (ii) test code, the agent prunes statements in the test code after the assertion failure.}
    \label{fig:context-agent}
\end{figure}

\phead{Generate failure-reason from code artifacts.} Our failure-reason generation makes the following three observations: (i) Firstly, the stack traces are verbose and include calls to external libraries unrelated to the fault. (ii) Secondly, some statements within a test case are irrelevant to the failure, specifically statements after the first assertion failures. (iii) Finally, the test case may call other helper methods that trigger the test failure~\cite{RevisitingTestImpact}. Hence, as shown in Figure~\ref{fig:context-agent}, to mitigate (i), the \textit{Context Extraction Agent} uses \texttt{PreprocessTrace} tool-chain to prune external execution in the stack trace. To mitigate (ii) and (iii), \textit{Context Extraction Agent} uses \texttt{PreprocessTest} tool-chain, which uses static analysis to build an interprocedural call graph to extract all helper test methods called by the test and prunes all the statements appearing in the test after the failure point.

Given the preprocessed stack trace and test code, we prompt \textit{Context-Extraction Agents} to generate \textit{failure-reason} by answering test purpose, expected output, and failure reason. Below, we show a simplified example of the prompt: 
\greybox{\textit{
``Analyze the test body and stack trace to generate a test failure reason summarizing the issue with expected output, and potential faulty location." 
        \\
        \#\# \textbf{Input}:
        `\{Test Code, Stack Trace\}'
        \\
        \#\# \textbf{Response}:
        `\{Failure-Reason\}'
        }}
        
An example of \textit{failure-reason} generated for \textit{Lang-5} given stack trace and test code is:
\greybox{\textit{
\#\# \textbf{Test Purpose}:
\\
Test whether LocaleUtils.toLocale parses '\_GB' as a locale string, expecting it to return a Locale object with an empty language and 'GB' as the country. 
\\
\#\# \textbf{Failure Reason}:
\\
The actual output is an IllegalArgumentException due to an invalid locale format, as LocaleUtils.toLocale requires a language before the underscore. For example, '\_GB' does not conform to the expected locale format. Update the test input to 'en\_GB' or modifying LocaleUtils.toLocale to handle '\_GB' as a valid case.
d}}

\phead{Failure-reason guided prioritization. }  
Given \textit{Failure-reason}, the \textit{Context Extraction Agent} analyzes the covered methods in each group, previously denoted as \( \{C_1, C_2, \dots, C_K\} \), to prioritize methods that are most related to the failure. Formally, prioritizing each set \( C_i \) results in a prioritized subset \( C'_i \), where: 
\( C'_i \subseteq C_i, \forall i \in \{1, 2, \dots, K\}\). The prioritized methods from all subsets are then combined
into a final prioritized set: \( C' = \bigcup_{i=1}^{K} C'_i\). This final prioritized list \( C' \) is just a union of the ordered subsets, meaning no additional processing is introduced. This union process is necessary as LLMs are highly sensitive to initial order~\cite{chen2024premise}, yet analyzing entire code coverage poses a token limitation. By dividing coverage into groups, prioritizing each, and unionizing subsets, we extract important context that are necessary for the \textit{Debugger Agent} for fault analysis and localization in the next step. 

\subsection{Debugger Agent}  
Previously, the \textit{Context Extraction Agent} prioritized \(C\) to subsets \( C' \) to mitigate the LLM's token limitation when analyzing large coverage data. While this technique can adapt to any LLM and token input size, prioritization may miss methods in the coverage that are important for fault localization. Inspired by how developers analyze the call dependencies to understand program execution and identify faults, we propose a \textbf{graph-based retrieval-augmented code navigation strategy (Graph-RAG)}.
This approach enables the \textit{Debugger Agent} to utilize call graphs and method bodies for effective repository navigation, implementation analysis, generation of \textit{failure reasoning}, and ranking the methods based on their likelihood of causing errors. We will detail this technique below.
\subsubsection{Generating failure reasoning through call-graph-aware retrieval-augmented (Graph-RAG) code navigation.} The agent first uses  code-coverage \( C \) to construct the inter-procedural call-graph database \( G = (V, E) \), where \( V \) is a set of methods in the call-graph, and \( E \) is a set of edges representing the caller-callee relationship. This graph is used by the tool-chains, \texttt{get\_MethodBody} and \texttt{get\_CallGraph} to facilitate agent-driven navigation within the call graph. \tool then uses the prompt template below to drive the agent: 
\greybox{\textit{
``Given (i) failure reasoning, (ii) stack trace, and (iii) test code, analyze, navigate, and enhance the failure reasoning for the given methods to identify faults. If additional implementation details are needed, extract the call graph and retrieve the caller or callee's implementation. At the end, you should output the (i) all of the analyzed method, (ii) enhanced failure reasoning, (iii) fault ranking." 
        \\
        \#\# \textbf{Input}: \\
        `\{Test Code, Stack Trace, Failure Reason, PrioritizedMethods\}'
        \\
        \#\# \textbf{Response}: \\
        `\{Analyzed Method, Enhanced Failure Reason, Fault Ranking\}'
        }}

The agent begins the analysis of methods selected sequentially from the prioritized \( C' \). For each method, the agent retrieves its body using \texttt{get\_MethodBody} and analyzes its implementation.
It then queries \texttt{get\_CallGraph} to identify relevant callers or callees, deciding whether to retrieve their implementations to enhance its understanding of code behavior. 
Throughout the iterative process of retrieval and analysis, every method examined, regardless of whether it was prioritized or invoked by a prioritized method, is assigned a failure reason by the agent. This information is then saved in a set referred to as \(R\).
This failure reasoning is a natural language explanation of each method's failure analyzed in the code navigation process. 
These results are stored as pairs in a set:
$R = \{ (m_i, r_i) \mid m_i \text{ is analyzed},\; i = 1, \dots, n \}$.
Here, \( n \) denotes the number of methods for which failure reasoning was generated. 


\phead{Ranking methods based on failure reasoning results.}
Finally, given the failure reasoning \(R\), the agent assigns each method \( m_i \in \mathcal{R} \) an ordinal rank (\( r^*_i \)). Formally, the final ranked set \(R^*\) is sorted in descending order based on the ordinal ranks:
\[
R^* = \text{sort}\left(\{(m_i, r_i, r^*_i) \mid m_i \in R\}\right)
\]
where \( r*_i\) denotes the ordinal rank that reflects each method's likelihood of containing the fault, and \( r_i \) explains the reasoning behind the method being faulty. The final ranked list \( R^* \) is output in JSON format for subsequent analysis.

\subsection{Reviewer Agent} Code review is important in software development to ensure software quality. Beyond traditional debugging, code review can help refine failure reasoning and improve fault localization. Inspired by this, we propose \textit{Reviewer Agent}, which uses a novel prompting technique called \textbf{Re-ranking through Verbal Reinforcement Learning} to emulate a rigorous code-review process to refine fault localization. We describe the technique below.


\subsubsection{Re-ranking through Verbal Reinforcement Learning}  
To refine the fault rankings \( R^* \), \textit{Reviewer Agent} adopts a verbal reinforcement learning process inspired by the Reflexion framework~\cite{shinn2024reflexion}. The agent iteratively critiques its rank trajectory by identifying inconsistencies, retrieving missing execution insights, and updating method prioritization to better reflect the failure’s root cause. After the iterations, it applies chain-of-thought to update the final ranking and return the fault localization results. 
\greybox{\textit{
``Given (i) failure reasoning, (ii) stack trace, and (iii) test code, analyze and critique the initial ranking of suspicious methods. Reflect on the reasoning quality and identify missing or superfluous details. If additional insights are needed, extract the call graph and retrieve the caller or callee's implementation to refine the ranking. At the end, you should output (i) the revised set of analyzed methods, (ii) the improved failure reasoning, and (iii) the final fault ranking with a possible fix."
        \\
        \#\# \textbf{Evaluating the Initial Ranking}: \\
        `\{Test Code, Stack Trace, Ranked List with Reasonings\}'
        \\
        \#\# \textbf{Self-Critique and Refinement}: \\
        - Identify missing insights and remove unnecessary information. \\
        - Retrieve method details using `get\_MethodBody` if needed. \\
        - Extract the call graph with `get\_CallGraph` to enhance fault reasoning. \\
        \\
        \#\# \textbf{Final Output}: \\
        `\{Analyzed Methods, Enhanced Failure Reasoning, Fault Ranking\}'
        }}

\phead{Evaluating and self-critiquing the initial ranking.}  
The Reviewer Agent uses the initial ranking, $R^*$, provided by the Debugger Agent as the baseline trajectory for refinement.
The agent enters a self-reflection phase, guided by Reflexion’s Self-Reflection Model (\(M_{sr}\)), to refine its ranking decisions through verbal reinforcement learning~\cite{shinn2024reflexion}.
In this phase, the agent examines whether the initial ranking aligns with observed fault rankings by comparing the failure reasoning \( r \), with caller-callee interactions and test coverage data, thereby identifying discrepancies such as misordered rankings or missing call dependencies. 
The agent generates natural language critiques and retrieves additional execution details if needed using tools such as \texttt{get\_MethodBody} (to obtain complete method implementations) and \texttt{get\_CallGraph} (to retrieve caller-callee relationships). 
The refined context is then incorporated in the Reviewer Agent's iterative adjustment process.

\phead{Ranking adjustment through trajectory optimization.}  
After the initial self-critique phase, the Reviewer Agent refines its initial ranking by updating each method's rank \( r^* \) and revising its failure reasoning \( r \). 
Conceptualized as a trajectory optimization problem within the Reflexion framework, the agent, acting as an Actor (\( M_a \)), leverages reinforcement cues from its Self-Reflection Model (\( M_{sr} \)) to systematically re-order methods based on the updated evidence of inter-method interactions and failure reasoning. Each iteration integrates new feedback to adjust the ranking until it stabilizes or a preset iteration limit is reached. 

\phead{Finalizing the ranking through chain-of-thought.} After the iterative ranking adjustment, the Reviewer Agent finalizes the ranking one last time using chain-of-thought~\cite{wang2024strategic}. To facilitate the thinking process, we ask the Reviewer Agent to generate a probable fix for every method in the final ranking by considering the updated ranking score \( R^* \) and refined failure reasoning \( R \). After generating all probable fixes, the agent then revisits all the information to do a final re-ranking. 
Formally, the final ranked list 
$R^*_{final} = \text{sort}\left(\{(m_i, r^*_i, r_i, f_i) \mid m_i \in C'\}\right)$, 
where \( f_i \) denotes the fix generated for method \( m_i \). 
The final ranked list \( R \), encoded in JSON format, provides a structured, machine-readable prioritization along with human-interpretable failure justifications. 

%% file: samples/results.tex
\section{STUDY DESIGN AND RESULTS}
\label{studyrq}
In this section, we first describe the study design and setup. Then, we present the answers to the research questions. 

\begin{table}
\caption{An overview of our studied projects from Defects4J v2.0.0. {\em \#Faults}, {\em LOC}, and {\em \#Tests} show the number of faults, lines of code, and tests in each system. {\em Fault-triggering Tests} shows the number of failing tests that trigger the fault. }
\scalebox{0.7}{
\setlength{\tabcolsep}{0.43cm}
\begin{tabular}{lrrrr}
    \toprule
    \textbf{Project} & \textbf{\#Faults} & \textbf{LOC} & \textbf{\#Tests} & \textbf{Fault-triggering Tests}\\
    
    \midrule
    Cli          & 39       & 4K     & 94     &  66      \\
    Closure      & 174      & 90K    & 7,911  &  545     \\
    Codec        & 18       & 7K     & 206    &  43      \\
    Collections  & 4        & 65K    & 1,286  &  4       \\
    Compress     & 47       & 9K     & 73     &  72      \\
    Csv          & 16       &  2K    & 54     &  24      \\
    Gson         & 18       & 14K    & 720    &  34      \\
    JacksonCore  & 26       & 22K    & 206    &  53      \\
    JacksonXml   & 6        & 9K     & 138    &  12      \\
    Jsoup        & 93       & 8K     & 139    &  144     \\
    Lang         & 64       & 22K    & 2,291  &  121     \\
    Math         & 106      & 85K    & 4,378  &  176     \\
    Mockito      & 38       & 11K    & 1,379  &  118     \\
    Time         & 26       & 28K    & 4,041  &  74      \\
    \midrule
    \textbf{Total}& 675 & 380K & 24,302 & 1,486 \\
    \bottomrule
\end{tabular}
\label{tab:overview}

}
\end{table}
\phead{Benchmark Dataset.}
To answer the RQs, we conducted the experiment on 675 faults across 14 projects from the Defects4J benchmark (V2.0.0)~\citep{just2014defects4j}. Defects4J provides a controlled environment to reproduce faults collected from various types and sizes of projects. Defects4J is widely used in prior automated fault localization research \cite{lou2021boosting, sohn2017fluccs, chen2022useful, zhang2017boosting}. 
We excluded three projects, JacksonDatabind, JxPath, and Chart, from Defects4J in our study because we encountered many execution errors and were not able to collect test coverage information for them. 
Table \ref{tab:overview} gives detailed information on the projects and faults we use in our study. The faults have over 1.4K fault-triggering tests (i.e., failing tests that cover the fault). The sizes of the studied projects range from 2K to 90K lines of code. Note that since a fault may have multiple fault-triggering tests, there are more fault-triggering tests than faults. 


\phead{Evaluation Metrics. }
According to prior findings, debugging faults at the class level lacks precision for effective location \cite{kochhar2016practitioners}. Alternatively, pinpointing them at the statement level might be overly detailed, omitting important context \cite{parnin2011automated}. Hence, in keeping with prior work \citep{benton2020effectiveness, b2016learning, li2019deepfl, lou2021boosting, vancsics2021call}, we perform our fault localization process at the method level. We apply the following commonly-used metrics for evaluation:

\uhead{\underline{Recall at Top-N}}. The Top-N metric measures the number of faults with at least one faulty program element (in this paper, methods) ranked in the top N. The result from \tool is a ranked list based on the suspiciousness score. Prior research \cite{parnin2011automated} indicates that developers typically only scrutinize a limited number of top-ranked faulty elements. Therefore, our study focuses on Top-N, where N is set to 1, 3, 5, and 10.

\phead{Implementation and Environment.} 
To collect test coverage data and compute results for baseline techniques, we utilized Gzoltar~\cite{campos2012gzoltar}, an automated tool that executes tests and gathers coverage information. For the LLM-based components, we employed OpenAI's gpt-4o-mini-2024-07-18, a more cost-effective yet capable LLM~\cite{openai_gpt4o_2024}. We used LangChain to develop \tool~\cite{langchain_docs_2024}. We designed the prompts to be concise to minimize token usage and to allow more room for analysis-related information and code. To reduce the variations in the output, we set the temperature parameter to 0 during model inference.

\input{samples/rq1}

\input{samples/rq2}

\input{samples/rq3}

%% file: samples/rq1.tex
\subsection*{RQ1: How does \tool perform compared with other fault localization techniques?}

\phead{Motivation.} In this RQ we evaluate the fault localization accuracy of \tool by comparing with various baseline techniques. 

\phead{Approach.} We compare \tool's fault localization accuracy with five baselines representing different methodological families: \textbf{\ochiai} (statistical)~\cite{abreu2006evaluation}, \textbf{\textit{DeepFL}} (deep neural network)~\cite{li2019deepfl}, {\bf \textit{Grace}} (graph neural network)~\cite{lou2021boosting}, \textit{\textbf{DepGraph} (graph neural network)}~\cite{depgraph}, \textit{\textbf{AutoFL} (LLM)}~\cite{autofl}, and \textit{\textbf{AgentFL/SoapFL}}~\cite{qin2024agentfl} (LLM).

\ochiai~\cite{abreu2006evaluation} is a widely recognized statistical fault localization technique known for its high efficiency, making it a common baseline for comparison~\cite{lou2021boosting, li2021fault, cui2020improving, wen2019historical, qian2021agfl, depgraph}. As such, we use \ochiai to rank the methods during the segmentation process and include it as a baseline for accuracy comparison.

\textit{DeepFL}~\cite{li2019deepfl} is a deep-learning-based fault localization technique that integrates spectrum-based and other metrics such as 
code complexity, and textual similarity features to locate faults. It utilizes a Multi-layer Perceptron (MLP) model to analyze these varied feature dimensions. We follow the study~\cite{li2019deepfl} to implement \textit{DeepFL} and include the SBFL scores from 34 techniques, code complexity, and textual similarities as part of the features for the deep learning model. 
\textit{Grace~\cite{lou2021boosting}} utilized graph neural networks (GNN). It represents code as a graph and uses a gated graph neural network to rank the faulty methods. 
\depgraph~\cite{depgraph} is another GNN-based technique that further improves \textit{Grace} by enhancing code representation in a graph using interprocedural call graph analysis for graph pruning and integrating historical code change information. 

\textit{AutoFL}~\cite{autofl} is an LLM-based fault localization approach that provides the LLM with a failing test and method descriptions to gather relevant coverage data, then repeats the process to assign inverse scores and rank candidate methods by averaging results across runs.
\textit{SoapFL} (also known as \textit{AgentFL})~\cite{qin2024agentfl} organizes fault localization into structured phases, including test failure comprehension, code navigation, and iterative LLM-based scoring, leveraging enhanced method documentation to refine suspicious methods.
While both \textit{AutoFL} and \textit{SoapFL} used OpenAI’s GPT-3.5 for their experiments, for our evaluation, we adapted both techniques to use the same LLM version as \tool (i.e., gpt-4o-mini-2024-07-18) for comparison.

\begin{table}
    \caption{\textbf{Fault localization accuracy (Top-1, 3, 5, and 10) for 675 faults from Defect4J V2.0.0. The numbers in the parenthesis show the percentage difference compared to LLM4FL.}}
    \centering
    \label{tab:rq1-2}
    \scalebox{0.8}{
    \setlength{\tabcolsep}{0.1cm}
    \begin{tabular}{l|llll}
    \toprule
        \textbf{Techniques} & \textbf{Top-1} & \textbf{Top-3} & \textbf{Top-5} & \textbf{Top-10} \\
        \midrule
        Ochiai & 121 (169.42\%) & 260 (63.46\%) & 340 (39.41\%) & 413 (20.10\%) \\
        DeepFL & 257 (26.85\%) & 353 (20.40\%) & 427 (11.01\%) & 468 (5.98\%) \\
        Grace & 298 (9.40\%) & 416 (2.16\%) & 486 (-2.47\%) & 541 (-8.32\%) \\
        DepGraph & 359 (-9.19\%) & 481 (-11.64\%) & 541 (-12.38\%) & 597 (-16.92\%) \\
        AutoFL & 275 (18.55\%) & 393 (8.14\%) & 423 (12.06\%) & 457 (8.53\%) \\
        SoapFL/AgentFL & 311 (4.82\%) & 414 (2.66\%) & 455 (4.18\%) & 478 (3.77\%) \\
        \midrule
        \tooltable & 326 & 425 & 474 & 496 \\
        
    \bottomrule
    \end{tabular}
    }
\end{table}

\phead{Results.} \textit{\textbf{\tool outperforms the LLM-based baselines, AutoFL and AgentFL, by achieving a 185.55\% and 4.82\% improvement in Pass@1, respectively.}}
Tables \ref{tab:rq1-2} show the fault localization results of \tool and the baseline techniques. Among the three LLM-based techniques, \tool achieves a better Top@N across all values of N. 
In the Top-1 metric, \tool locates the correct fault in 326 cases, compared to \textit{AutoFL}’s 275 and \textit{AgentFL}'s 311, representing an 18.55\% and 4.82\% improvement, respectively. Similarly, in Top-3, Top-5, and Top-10, \tool an improvement between 8.14\% to 12.06\% over AutoFL and 2.66\% to 4.18\% for \textit{AgentFL}. 
These numbers highlight \tool's ability to pinpoint faulty methods more accurately. For cost, we compute the total dollars spent by multiplying each approach’s token usage by the per-million-token price (e.g., \$0.150 for input, \$0.600 for output). \textit{\ul{SOAPFL averages about \$0.055 per bug, \textit{AutoFL} averages about \$0.065, while LLM4FL is at around \$0.050 per bug making our approach comparably cost-effective}}.


\noindent\textit{\textbf{\tool shows higher Top-1 and Top-3 compared to most other non-LLM-based techniques.}} 
For the Top-1 metric, \tool scores 326, which is 169.42\% higher than Ochia's score of 121, 26.84\% higher than DeepFL’s score of 257, and 9.39\% better than Grace’s result of 298. 
One exception is \depgraph, which achieves a Top-1 of 359, 8.64\% higher than \tool. 
As the range expands to Top-3 and beyond, \tool demonstrates its robustness, significantly outperforming DeepFL and maintaining strong performance alongside Grace. 
LLM-based techniques have several advantages over traditional techniques such as \depgraph. First, techniques like \tool leverage pre-trained LLMs in a zero-shot setting, which can be easily applied to systems with insufficient training data. Second, techniques like \depgraph require additional model training that can take days. LLM-based techniques leverage generic pre-trained LLMs without specific fine-tuning or re-training. Finally, LLM-based techniques can explain the decision, which helps with adaption~\cite{autofl, qin2024agentfl}. Hence, LLM-based techniques have a strong potential to enhance fault localization by offering greater adaptability, reducing training overhead, and providing interpretable explanations.

\rqboxc{\tool achieves a higher Top-1 compared to other LLM-based techniques, AutoFL and AgentFL, by 18.55\% and 4.82\%, respectively. It also has competitive results compared to supervised techniques like DeepFL and Grace by leveraging a pre-trained LLM in a zero-shot setting.  
}

%% file: samples/rq2.tex
\subsection*{RQ2: How do different components in \tool affect the fault localization accuracy?} 

\phead{Motivation.} \tool employs a number of components, and 
each of these components plays a distinct role in the overall process. Hence, in this RQ, we conduct an ablation study by removing each component separately and study their impact on fault localization accuracy. 
The findings may inspire future studies on adapting the components for similar tasks. 

\phead{Approach.} To evaluate the impact of each component, we designed four different configurations: 

\uhead{\textbf{\toolwopromptchain}} removes the code navigation mechanism, which essentially means the Debugger agent no longer does fault navigation by retrieving the call graphs. Instead, the \tool uses a single prompt to perform fault localization without fault navigation. This configuration tests whether using the caller-callee information improves the ranking and selection of faulty methods or if a single round of analysis is sufficient.

\uhead{\textbf{\toolwosplitting}} removes the order-aware division of the covered methods, giving the agents the entire coverage data at once instead of dividing it into smaller, manageable groups. Coverage segmentation addresses token limitations in LLMs, so removing it explores the impact of feeding the full dataset to the agents in one step. We aim to see how handling large amounts of data in a single input influences the fault localization result, as it may overwhelm the model or reduce precision.

\uhead{\textbf{\toolworeflection}} removes the verbal reinforcement learning technique, which is used to allow agents to review and refine their initial ranking. Without this step, the agents rely solely on their initial assessments without iterative improvements. 

\begin{table}
    \caption{Impacts of removing different components in \tool on Top-1, 3, 5, and 10. The numbers in the parentheses show the percentage changes compared to \tool with all the components.}
    \vspace{-0.3cm}
    \centering
    \label{tab:rq4}
    \scalebox{0.8}{
    \setlength{\tabcolsep}{0.09cm}
    \begin{tabular}{l|cccc}
    \toprule
        \textbf{Techniques} & \textbf{Top-1} & \textbf{Top-3} & \textbf{Top-5} & \textbf{Top-10} \\
        \midrule
        \tooltable & 327 & 425 & 473 & 494 \\
        \toolwopromptchaintable & 273 (-16.51\%) & 378 (-11.06\%) & 409 (-13.53\%) & 409 (-17.21\%) \\
        \toolwosplittingtable & 251 (-23.24\%) & 341 (-19.76\%) & 365 (-22.83\%) & 381 (-22.87\%) \\
        \toolworeflectiontable & 290 (-11.31\%) & 400 (-5.88\%) & 436 (-7.82\%) & 459 (-7.09\%) \\
    \bottomrule
    \end{tabular}
    }
    \vspace{-0.3cm}
\end{table}

\phead{Results.} \textbf{\textit{While all components help improve the results, including coverage division and code navigation provide the largest improvement to fault localization results (23\% and 17\% in Top-1).}} Table~\ref{tab:rq4} shows the Top-1, 3, 5, and 10 scores when each component is removed. Removing coverage divisions has the largest overall impact across all scores, reducing Top-N by 19\% to 23\%. Removing prompt chaining has the second largest impact (11\% to 17\%). 
At the individual project level, these two components also have the largest impact in Top-1 in 9/13 studied projects.
Our finding shows that employing sorted coverage grouping following the divide and conquer technique and agent communication significantly improves fault localization results. Future research should consider these techniques when designing fault localization techniques. 

\noindent \textit{\textbf{Although there is no oracle during the fault localization process, asking LLMs to self-reflect still helps improve the overall Top-1 by 11\%.}} LLMs often suffer from hallucinations, especially when there is a lack of feedback from external oracles~\cite{xu2024hallucination, huang2023survey}. Even though we did not provide any groundtruth or external feedback to LLM, we found that reflexion is still effective in improving fault localization results. 
We speculate that verbal reinforcement learning helps the model improve its results by revisiting the suspicious methods it ranked earlier, creating a feedback loop. In this process, we analyze the results for each group of methods and combine these group-specific results into a final ranking. The model can then spot mistakes or gaps in its logic, leading to better results.
Self-reflection brings 6\% to 11\% improvement across the Top-N metrics. Our finding suggests that future studies should consider self-reflection even if there is no external feedback. 
Our finding highlights the effectiveness of self-reflection, which should be considered in future fault localization results.

\rqboxc{The results show that each component of \tool contributes to its overall fault localization performance, with coverage division and code navigation having the largest positive impact. Removing these components leads to significant declines in accuracy, confirming their role in finding faults in a large code-base.
}

%% file: samples/rq3.tex
\subsection*{RQ3: Does order matter in the initial list of methods provided to the LLM?}

\phead{Motivation.} 
\tool divides the coverage data into different divisions using an order-aware division strategy to address the token size limitation of LLMs. We sort the methods using the \ochiai scores before the division, though different sorting mechanisms may affect the final fault localization result. 
Although \tool eventually visits and assesses every method, a recent study~\cite{chen2024premise} observes that the order of premises affects LLM's results. However, whether this effect extends to software engineering tasks, particularly fault localization, remains unclear. 
Hence, in this research question, we investigate whether the order of methods within the groups affects the LLM's fault localization performance. 
 

\phead{Approach.} To test the effect of method ordering, we experiment with three distinct sorting strategies: \textbf{\toolrandom}, \textbf{\toolochiai} (the default sorting in \tool), and \textbf{\tooldepgraph} to sort the methods before the divide-and-conquer step. 

\uhead{\textbf{\toolrandom:}} We use the unsorted list of methods executed during testing, as generated by Gzoltar~\cite{campos2012gzoltar}. This default list represents the natural execution order of the methods, with no explicit ranking or prioritization. By providing the LLM with methods based on the execution sequence, we establish a control case to measure its performance without any ranking influence. 

\uhead{\textbf{\toolochiai}:} As discussed in Section~\ref{sec:splitting}, we apply \ochiai to sort the methods before the divide-and-conquer process. \ochiai is unsupervised and is efficient to compute. We hypothesize that providing the LLM with methods sorted by their suspiciousness score will lead to more effective fault localization, as the model will focus on the most likely faulty candidates earlier in the process.

\uhead{\textbf{\tooldepgraph}:} This approach uses the ranking produced by \depgraph, a state-of-the-art Graph Neural Network (GNN)-based fault localization technique~\cite{li2015gated,depgraph}, to sort the methods. \depgraph ranks methods based on structural code dependencies and code change history. 
As shown in RQ1, \depgraph shows the highest fault localization accuracy among all the techniques, surpassing \toolochiai. By examining the fault localization result after sorting the methods using \depgraph's scores, we can better study if the initial order affects LLM's results, even though LLM eventually visits all the methods. 



\begin{figure}
    \centering
    
    \includegraphics[width=\linewidth]
    {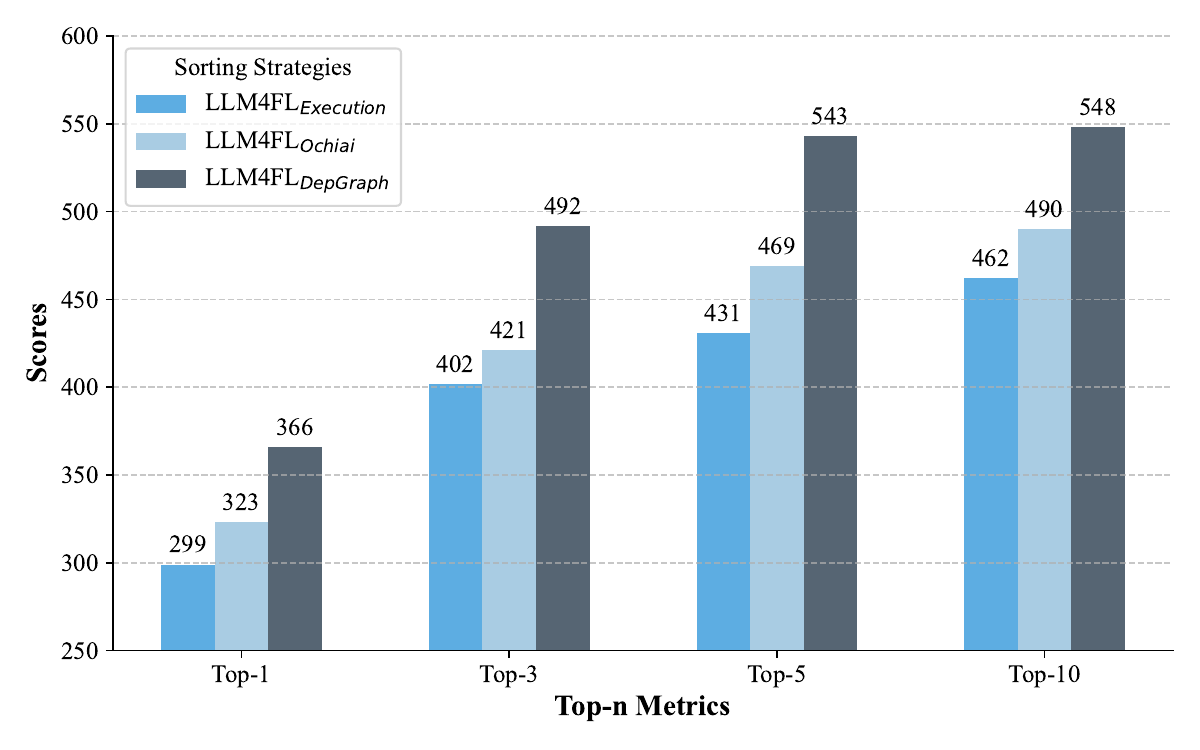}
    \caption{Fault localization results when using different method sorting strategies during the segmentation process. }
    \vspace{-0.3mm}
    \label{fig:rq2}
\end{figure}
\phead{Results.} \textbf{\textit{Method ordering has a significant impact 
on LLM's fault localization result, with up to 22\% difference in Top-1 (from 299 to 366).}} Figure~\ref{fig:rq2} shows the fault localization results when using different sorting strategies. 
When methods were presented in the execution order, \textit{\toolrandomtable} achieved a Top-1 score of 299, 402 for Top-3, 431 for Top-5, and 462 for Top-10. This performance establishes a baseline, showing how the LLM behaves without strategic ordering. However, sorting methods with the lightweight \textit{\ochiai} scores resulted in noticeable improvements across all Top-N, where \textit{\toolochiaitable} improved the Top-1 score to 323, an 8\% increase over \toolrandom. 

\noindent\textbf{\textit{\tooldepgraph provides further improvement to the already-promising result of \depgraph, indicating method ordering is critical to \tool, or LLM-based fault localization in general.}}
\textit{\tooldepgraphtable} achieved the highest Top-1 score of 366, which significantly outperforms both \toolrandomtable and \toolochiaitable. The improvement was consistent across all the metrics. 
We also see that \tooldepgraph has better Top-1, 3, and 5 scores compared to \depgraph. 
This consistent improvement underscores the importance of method ordering in enhancing the accuracy of LLM-based fault localization. 
Namely, if the initial order is closer to the group truth, the final localization result also tends to be more accurate. 

Our finding establishes a new research direction for LLM-based fault localization, or any software engineering tasks that take a list of software artifacts as input. Future studies may study how different premises of ordering affect other software engineering tasks, and how to combine traditional software engineering techniques to pre-process LLM's input to further improve the results. 

\rqboxc{The initial method ordering significantly impacts the accuracy of LLM-based fault localization, with Top-1 scores varying by up to 22\%. Future research should explore different ordering strategies and how traditional software engineering techniques can be integrated to optimize LLM performance further.} 

%% file: samples/threats.tex
\section{Threats to Validity}
\label{threats}

\phead{Internal Validity.}  
A potential threat to internal validity is the risk of data leakage in large language models (LLMs), where the model might have been exposed to the benchmark data during training. 
Nevertheless, Ramos et al.~\cite{ramos2024large} found that newer and larger models trained on larger datasets exhibit limited evidence of leakage for defect benchmarks. Since we utilize GPT-4o-mini, a large model trained using a tremendous amount of data, it likely shares similar characteristics in reducing memorization risks. Additionally, following prior work, we ensure no content related to the project name, human-written bug report, or bug ID is entered into ChatGPT to minimize the risk of data memorization~\cite{qin2024agentfl}.

\phead{External Validity.}
The primary threat to external validity is the generalizability of our results. Our evaluation is based on Defects4J, a well-established dataset in the software engineering community. Although this dataset includes real-world bugs, the systems studied are primarily Java-based. Future studies may extend our study to other programming languages or domains. 

\phead{Construct Validity.}
Construct validity relates to whether the metrics we used accurately measure the performance of fault localization techniques. We used widely accepted Top-N metrics commonly utilized in prior fault localization studies. However, our results are based on the assumption that developers primarily focus on the top-ranked faulty methods. Although this assumption aligns with previous research, different development practices could influence the effectiveness of our approach.

%% file: samples/conclusion.tex
\section{Conclusion}
\label{conclusion}
In this paper, we introduced \tool, an LLM-agent-based fault localization approach. It utilizes multiple specialized LLM agents, including Context, Debugger, and Reviewer, to iteratively refine and improve the accuracy of fault localization through order-aware prioritization, graph-based code navigation, and self-reflection using verbal reinforcement learning. Evaluated on the Defects4J (V2.0.0) benchmark, \tool demonstrated significant improvements over existing approaches, achieving an 18.55\% increase in Top-1 accuracy compared to AutoFL and 4.82\% improvement over SoapFL. Further enhancements, including coverage segmentation and iterative refinement, increased accuracy by up to 22\%. Future work will explore expanding \tool's capabilities for larger and more diverse codebases, further refining the agent collaboration and reasoning mechanisms.